# Topological sensing with photonic arrays of resonant circular waveguides

Kiernan E. Arledge,[1] Bruno Uchoa,[2] Yi Zou,[3] and Binbin Weng[1,*]

[1]*School of Electrical and Computer Engineering, University of Oklahoma, Norman, Oklahoma 73019, USA*
[2]*Center for Quantum Research and Technology, Department of Physics and Astronomy,*
*University of Oklahoma, Norman, Oklahoma 73019, USA*
[3]*School of Information Science and Technology, ShanghaiTech University, Shanghai 201210, China*

We propose a photonic array of resonant circular dielectric waveguides with subwavelength grating that can be designed as a robust and sensitive topological chemical sensor. The device can detect trace amounts of a given chemical species through photonic edge modes that are impervious to most sources of disorder. We perform a simulation in the midinfrared that accounts for the absorption loss introduced by chemical molecules in contact with a strongly coupled photonic lattice of resonators. Due to the topological nature of the device, its chemical sensitivity scales linearly with the system size and can reach parts-per-billion range at the millimeter scale. Our findings suggest that topological chemical sensors could empower the development of novel on-chip integrated photonic sensing technologies.

## I. INTRODUCTION

The ability to reliably detect trace amounts of chemical molecules through integrated circuits is an important research and technological challenge in various fields, from carbon capture and environmental monitoring to medicine [1–4]. On-chip sensors based on infrared photonic approaches, e.g., tunable diode laser absorption spectroscopy (TDLAS), permit targeting specific chemical species by tuning light to resonate at the absorption frequency of the molecules at extremely low trace amounts [5]. Besides having excellent chemical selectivity and sensitivity, they also have a fast response/recovery rate, low environmental dependence and low power consumption [6,7] and thus are regarded as promising for applications that demand state-of-the-art chemical sensing. However, fundamental limitations posed by fabrication resolution and intrinsic disorder remain challenging barriers to overcome. For instance, optical scattering and cross-talking effects due to fabrication disorder can lead to significant transmission loss [8]; in optical resonators, size errors introduced by fluctuations in the manufacturing process detune the targeted frequency from the rest of the spectrum, thus impeding coherent transport in resonance-based optical sensors [9].

Since its inception with the quantum Hall effect, the concept of topological protection has raised the possibility of novel devices that are insensitive to disorder and structural perturbations. Attempts to extend topological concepts to photonics were first realized on gyromagnetic photonic crystals operating in the microwave regime [10–12]. Subsequent efforts were focused on extending topological photonic states with no external magnetic field [12–15]. Hafezi *et al.* [16] experimentally realized a two-dimensional aperiodic lattice of weakly coupled optical resonators. A tight-binding analysis predicted topologically protected chiral edge states analogous to the integer quantum Hall effect. In the strong coupling regime, it has been shown [17] that this system can form two time-reversed copies of the anomalous quantum Hall effect [18], thus behaving as a topological insulator [19], with propagating and counter propagating edge modes. Edge states have been successfully employed as sources of quantum light [20] and as topological lasers [21].

In this paper, we propose a conceptually novel topological sensing framework based on absorption spectroscopy using photonic arrays of strongly coupled circular waveguides, illustrated in Fig. 1. The topological protection ensures the robustness of the optical modes traveling at the edge to the aforementioned disorder effects. We identify the phase diagram where the topological photonic states are robust against the energy loss induced by a concentration of a chemical species. Using a subwavelenth grating (SWG) design that enhances interactions between light and matter at the waveguides [see Fig. 1(a)], we perform a midinfrared simulation of the energy loss in the photonic lattice in the presence of a trace concentration of methane ($CH_4$). Due to its topologically protected nature, the detection limit scales inversely with the system size and can be improved by several orders of magnitude by scaling up the lattice. In particular, we find that sensitivity in the parts per billion (ppb) level can be reached in photonic lattices at the millimeter scale. We propose that this topologically enabled design strategy provides a new pathway for developing integrated photonic sensing circuits. We also suggest the feasibility of such topological design for gas sensing applications.

*binbinweng@ou.edu

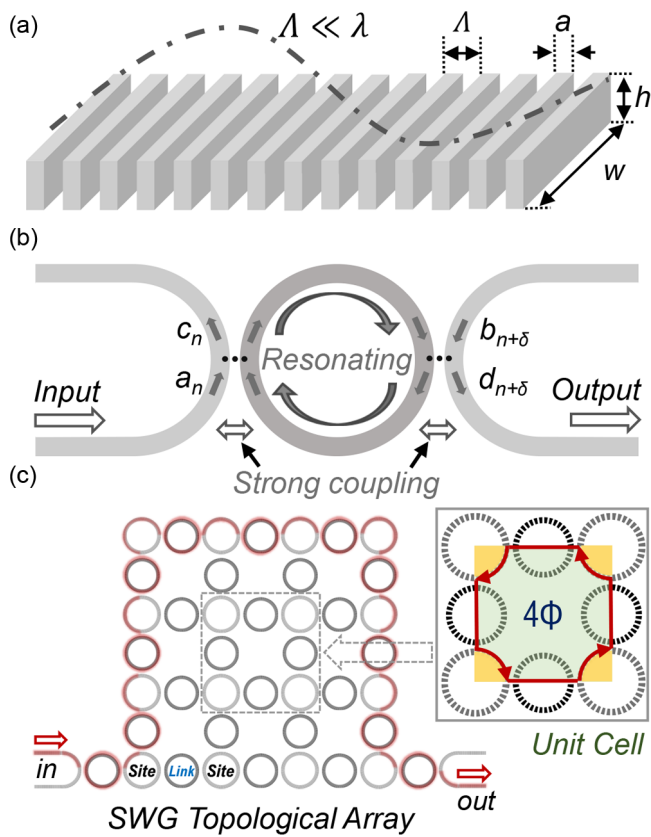

FIG. 1. (a) Subwavelength grating (SWG) waveguide for the enhancement of light-chemical interaction. (b) Strong coupling bridge resonator unit, which mediates the coupling between lattice sites in the array. (c) Design of the dielectric circular array with a source and drain of transmitting power. Gray rings: lattice sites. Black rings: bridge resonators connecting lattice sites. All rings are SWG waveguides, as shown in the inset. The red lines indicate the topologically protected modes propagating along the edges of the array. The inset shows a unit cell of the lattice. Light picks a phase of $\phi$ as it travels a quarter of a site ring, resulting in a geometric phase of $4\phi$ as it undergoes a closed circuit around the unit cell (shown as the red path). Green area: synthetic magnetic flux due to the geometric phase. Yellow filling: opposite synthetic flux. The net synthetic flux in the unit cell for each pseudospin is zero, as in the Haldane model.

The structure of the paper is as follows. In Sec. II, we describe the photonic lattice and the incorporation of absorption loss due to the presence of a chemical species. In Sec. III, we calculate the photonic bands based on the scattering matrix formalism and address the topological phase diagram as a function of absorption loss. In Sec. IV, we perform an absorption spectroscopy simulation in the midinfrared using realistic parameters for the absorption lines of methane and perform a scaling analysis of the simulation results for large system sizes. In the Appendix, we demonstrate that the proposed topological design is robust in the presence of intrinsic disorders, which can appear in the form of lattice dislocations and fabrication resolution effects.

## II. PHOTONIC LATTICE

We model a topological photonic insulator after Ref. [17] by constructing a square lattice of optical resonators, as shown in Fig. 1(c). The resonators each have the form of circular waveguides (gray circles) made of a dielectric material. Every lattice site is connected to a nearest neighbor (NN) site by a bridge resonator (black circles). For a given frequency, all resonators have one propagating and one counterpropagating optical mode, which operate as a two component pseudospin. In the absence of backscattering, the two pseudospins can be treated independently and their indexes will be omitted.

The scattering matrix relates the transmission of light between two NN resonators with incoming waves that have complex amplitudes $a_n$ and $b_{n+\delta}$, respectively, and outgoing waves $c_n$ and $d_{n+\delta}$, namely,

$$S_{n,x}\begin{pmatrix} a_n \\ b_{n+x} \end{pmatrix} = \begin{pmatrix} d_{n+x} \\ c_n \end{pmatrix}, \quad (1)$$
$$S_{n,y}\begin{pmatrix} d_n \\ c_{n+y} \end{pmatrix} = \begin{pmatrix} b_{n+y} \\ a_n \end{pmatrix} e^{-2i\phi},$$

as shown in Fig. 1(b). $S_{n,\delta}$ is the scattering matrix along the NN direction $\delta = x, y$.

The parameter $\phi$ in Eq. (1) is the phase that light acquires as it travels across a quarter of a site ring [17]. The inset of Fig. 1(c) depicts the counterclockwise round trip circuit that light travels in one unit cell of the resonator lattice. As light circulates around the unit cell, it picks up a $4\phi$ geometric phase. That phase corresponds to a finite synthetic magnetic flux in the area inside the optical path (green region), in analogy with the quantum Hall effect for electrons [16], and the opposite synthetic flux in the yellow region outside the area contained by the optical path. The net synthetic flux in the unit cell is therefore zero, resulting in one chiral mode per edge, as in the Haldane model [17,18]. The other pseudospin mode of light corresponds to a clockwise round trip around the unit cell, with the opposite geometric phase. The two pseudospins thus describe two copies of the anomalous quantum Hall effect, each one with opposite chiral edge modes. This construction is hence analogous to the quantum spin Hall effect [19], with two counter propagating modes per edge.

In a real ring resonator, $\phi$ depends on the light wavelength $\lambda$, effective refractive index $n$, and circumference of the ring $L_s$ as $\phi(\lambda) = \pi n L_s/2\lambda$ [22,23], and plays the role of the frequency in the photonic band structure. The scattering matrix can be parametrized to accommodate the lossy coupling by including exponential decay terms, which represent the effect of absorption loss on the wave amplitudes,

$$S_{n,\delta} = \begin{bmatrix} r_{n,\delta} & t'_{n,\delta} \\ t_{n,\delta} & r'_{n,\delta} \end{bmatrix} = \begin{bmatrix} \sin\theta e^{-\alpha L_{NN,\delta}/2} & i\cos\theta \\ i\cos\theta & \sin\theta e^{-\alpha L_{NN,\delta}/2} \end{bmatrix}. \quad (2)$$

The parameters $r$ and $t$ are the complex reflection and transmission coefficients, respectively, with $\theta \in [0, \frac{\pi}{2}]$ the phase that describes the coupling strength between NN sites. $\theta \sim 0$ describes the weak coupling regime, where the sites are nearly decoupled, whereas $\theta \sim \frac{\pi}{2}$ corresponds to the strong coupling limit ($t \approx 1$). The quantity $\alpha L_{NN,\delta}$ gives the energy loss due to absorption of light due to contact with a chemical species as it travels between two NN sites along the $\delta = x, y$ direction.

Here, $\alpha$ is the absorption loss per unit length due to the attenuation and $L_{\text{NN},\delta}$ is the optical path length of the light traveling between two NN sites mediated by a bridge resonator.

We design the lattice in a way that light with wavelength $\lambda$ travels multiple times around the bridge resonators with refractive index $n$. The optical path length in the bridge resonators $L_b$ is defined by the quality factor ($Q$ factor), which is a measure of the resonant photon lifetime and is thus connected to the effective path length light travels in the bridge resonators, $L_b = Q\lambda/2\pi n$. The resonance condition is that the effective light wavelength is a submultiple of the perimeter of the circular waveguide, $\lambda m = 2\pi R_b n$, with $R_b$ the radius of the bridge ring and $m$ is an integer. Hence,

$$L_b = Q R_b / m. \quad (3)$$

The quality factor $Q$ is material and geometry dependent. Typical $Q$ values for chemical sensing are in the range $10^4$–$10^8$, whereas $m$ is generally of the order of 10–100. Hence, $L_b$ is several orders of magnitude larger than the circumference of the bridge resonators in the high-$Q$ regime. At the same time, we design the optical path length of the lattice sites to be about the circumference of the rings $L_s$, such that $L_b/L_s \sim 10^2$–$10^6$. Therefore the sensitivity of the photonic lattice to molecules in air is determined by processes in which the gas mostly absorbs energy from the photons confined in the bridge resonators ($L_b \gg L_s$). In that case, to a very good approximation, we can assume that $L_{\text{NN},\delta} \approx L_b$ in Eq. (2).

In solid dielectric waveguides, light-matter coupling occurs through evanescent waves at the surface of the material, which are typically weak. We use instead subwavelength grating (SWG) waveguides. The grating enhances light-matter interactions by permitting the guided mode to overlap with the gas, which diffuses freely through the grating. The structure is characterized by a lattice period $\Lambda$, as depicted in Fig. 1(a). At operation wavelengths $\lambda \gg \Lambda$, one can treat the SWG structure as a homogeneous waveguide with a tunable effective refractive index set by geometry [24,25].

The optical power attenuation $T$ in the waveguide of the resonators decays exponentially with the gas concentration $\sigma$, the molar absorption $\epsilon$, and the optical path length $L_b$, namely $T = e^{-\gamma \epsilon \sigma L_b}$. The geometric fill factor $\gamma = (\int_V [\Re(n)]^2 |E|^2 d\mathbf{r})_{\text{air}}/(\int_V [\Re(n)]^2 |E|^2 d\mathbf{r})_{\text{total}}$ quantifies how much field energy overlaps with the analyte [26], with $E$ the position dependent electric field of the waveguide mode. From Eq. (3), the field attenuation that enters in the transfer matrix (2) has the form

$$\alpha L_b = \gamma \epsilon \sigma \frac{Q R_b}{2m}. \quad (4)$$

Solid ring resonators have a relatively small geometric factor of $\gamma \sim 0.1$–$0.2$ [8,27,28]. In SWG waveguides, optimized values range from $\gamma \sim 0.34$–$0.55$.

## III. TOPOLOGICAL PHASE DIAGRAM

For a suitable range of parameters, each pseudospin mode propagates independently both along the top edge of the lattice [shown in red in Fig. 1(c)] and along the bottom edge of the lattice. In the absence of processes that mix the two pseudospin flavors (backscattering) or else energy losses that are strong enough to close the topological gap, those edge modes are generically protected. Unlike in their electronic counterpart (topological insulators), the number of photons is not conserved and hence energy losses may destroy topological states even in the absence of backscattering. While the analogy with electronic systems is useful, there are significant differences between photonic topological systems and topological insulators regarding the physical mechanisms that protect edge modes against generic perturbations [30].

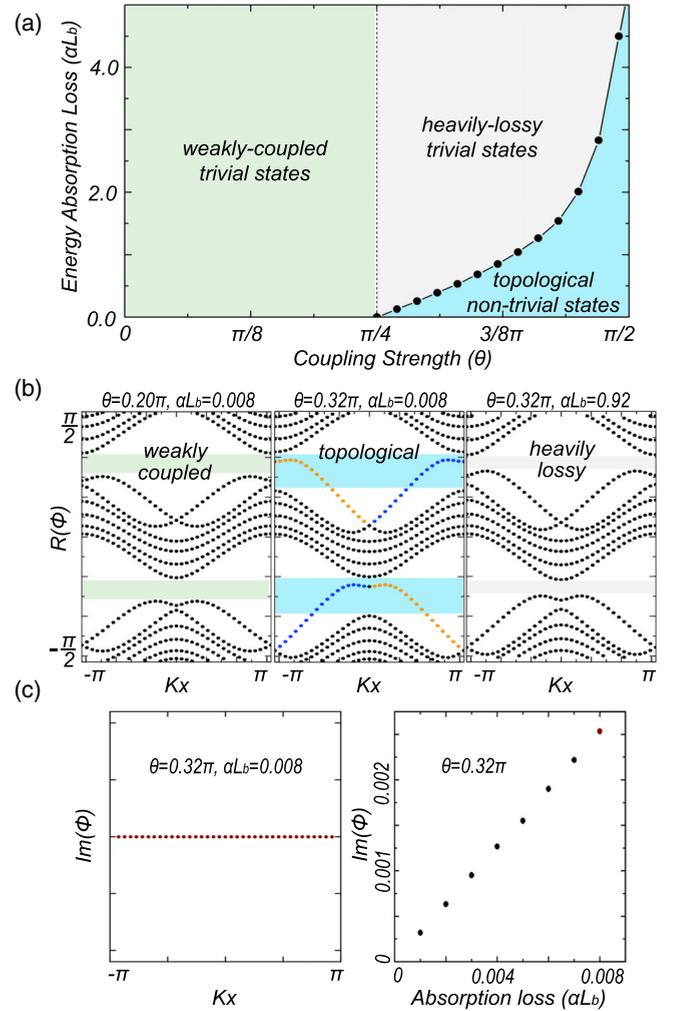

FIG. 2. (a) Phase diagram of energy loss $\alpha L_b$ vs coupling strength $\theta$ for a finite lattice ($N = 6$). The light blue region indicates a topological phase, in the strong coupling regime. (b) Projected bands $\Re[\phi(\lambda)]$ along $k_x$ in the three different phases. (Left) Trivial bands in the weakly coupled regime with $\theta = \pi/5$ and $\alpha L_b = 0.008$. (Middle) Topological edge states in the strong coupling regime for $\theta = 0.32\pi$ and $\alpha L_b = 0.008$. Orange and blue dots: topological edge states. Blue area: topological band gap. (Right) Heavily lossy phase ($\theta = 0.32\pi$ and $\alpha L_b = 0.92$), where the photonic modes again become topologically trivial. (c) $\Im[\phi(k)]$ versus momentum $k_x$ (left) and absorption loss (right) for $\theta = 0.32\pi$. $\Im[\phi(k)]$ scales linearly with $\alpha L_b$.

The topological phase diagram for energy loss ($\alpha L$) versus coupling strength $\theta$ can be extracted from the photonic band structure [see Fig. 2(a)]. Assuming a semi-infinite strip of ring resonators with $N$ unit cells along the $y$ direction (and periodic boundary conditions along the $x$ direction), one can calculate the projected photonic energy bands using the scattering matrix formalism. Since light picks a nontrivial phase $\phi$ only when it travels along a site ring, rather than across bridge sites, this photonic lattice model is not amenable to a tight-binding representation. This is an important distinction with other proposals for the simulation of the quantum Hall effect with photonic lattices, where the nontrivial topological phase is picked through "hopping" processes between sites.

We start with Eq. (1), which relates the amplitudes in site $N$ with its NN in the $x$ and $y$ directions. Since the lattice is assumed to be periodic in the $x$ direction, it is convenient to rewrite Eq. (1) in that direction in the Bloch form:

$$S_{n,x}\begin{pmatrix} a_n \\ b_n e^{ik_x} \end{pmatrix} = \begin{pmatrix} d_n e^{ik_x} \\ c_n \end{pmatrix}, \quad (5)$$

with $k_x$ the momentum along the $x$ direction written in dimensionless units. Reordering the scattering matrices in both directions, one obtains

$$\begin{aligned} S'_{n,x}(k_x)\begin{pmatrix} b_n \\ a_n \end{pmatrix} &= \begin{pmatrix} c_n \\ d_n \end{pmatrix}, \\ S'_{n,y}\begin{pmatrix} d_n \\ c_{n+y} \end{pmatrix} &= \begin{pmatrix} a_n \\ b_{n+y} \end{pmatrix}e^{-2i\phi}, \end{aligned} \quad (6)$$

where the new scattering matrices $S'_x$ and $S'_y$ now take the form

$$\begin{aligned} S'_{n,x}(k_x) &= \begin{bmatrix} r'_{n,\delta}\, e^{ik_x} & t_{n,\delta} \\ t'_{n,\delta} & r_{n,\delta}\, e^{-ik_x} \end{bmatrix}, \\ S'_{n,y} &= \begin{bmatrix} t_{n,\delta} & r'_{n,\delta} \\ r_{n,\delta} & t'_{n,\delta} \end{bmatrix}. \end{aligned} \quad (7)$$

Setting the boundary conditions at the edge of the lattice to be the same as the lossless case, namely, $c_1 = e^{-2i\phi}b_1$ and $a_N = e^{2i\phi}d_N$ [17,29], we calculate the projected band diagrams using the scattering matrices $S'_x$ and $S'_y$. As in the standard scattering matrix formalism, one can build larger $2N \times 2N$ matrices $M_A$ and $M_B$ for scattering processes along the $x$ and $y$ directions of the photonic lattice, respectively,

$$M_A = \begin{bmatrix} 1 & & & & \\ & S'_{1,y} & & & \\ & & \ddots & & \\ & & & S'_{N-1,y} & \\ & & & & 1 \end{bmatrix},$$

$$M_B(k_x) = \begin{bmatrix} S'_{1,x} & & \\ & \ddots & \\ & & S'_{N,x} \end{bmatrix}. \quad (8)$$

If we represent field amplitudes in the lattice site $n$ by the two component vector $\psi_n = (b_n, a_n)^T$, we can impose that the matrix product $M_A M_B(k_x)$ satisfies the eigenvalue equation $M_A M_B(k_x)\Psi = e^{-2i\phi}\Psi$, where $\Psi = (\psi_1, \ldots, \psi_N)^T$. To derive the bandstructure of the other pseudospin, one reverses the direction of propagation of the photon in the scattering matrix between NN sites (for fixed $\phi$) and inverts the boundary conditions at the edges. That procedure results in a degenerate spectrum, with the edge modes for the opposite pseudospin having their chirality reversed.

Adsorption processes do not by themselves produce backscattering. This formalism incorporating energy loss accounts for a single pseudospin at a time and explicitly ignores backscattering effects, which promote mixing between different pseudospins. It should be noted that many factors that cause energy loss in photonic structures (such as defects and dislocations of lattice sites) may in principle produce backscattering (they will be explicitly addressed later on). Recent experimental studies have shown that common intrinsic disorder effects do not lead to observable backscattering effects in the regime of strong coupling between sites [30]. The following calculation addresses how *absorption* losses suppress topological protection in the absence of backscattering effects. A detailed simulation incorporating backscattering will be considered in the next section, where we perform a realistic simulation of gas sensing via absorption spectroscopy accounting for defects.

The scattering matrices which describe lossy couplings are nonunitary and hence the eigenvalue equation derived above will result in the quasienergy $\phi$ being generally complex in nature and thus results in a complex band diagram. By solving this eigenvalue equation for each value of $k_x$, we can plot the real and imaginary components of the projected band diagram $\phi(k)$ along the $x$ direction. The real components $\Re[\phi(k)]$ are shown in Fig. 2(b). Those bands have an imaginary part, $\Im[\phi(k)]$, which is related to the energy loss of the modes due to adsorption.

For a given value of energy loss $\alpha L_b$, topological states are observed if the NN coupling $\theta$ is sufficiently strong. Figure 2(a) shows a phase diagram of energy loss $\alpha L_b$ versus coupling strength $\theta$. The critical solid line separating the different regions indicates a topological phase transition between states with topologically nontrivial edge modes (blue region) and topologically trivial ones (gray and green). At zero energy loss ($\alpha L_b = 0$), the critical coupling is $\theta_c = \pi/4$ [17]. In the strong coupling regime $\theta > \pi/4$ and $\alpha L_b = 0$, the system opens a topological gap, with two counter-propagating modes per edge, one for each pseudospin. Absorption losses effectively weaken the coupling between lattice sites. When strong enough, those losses eventually close the topological band gap, making the array topologically trivial.

In Fig. 2(b), we calculate the energy bands projected along the $k_x$ direction for a strip with $N = 6$ resonators in the $y$ direction. The left panel of Fig. 2(b) shows the band diagram of a lattice operating in the weak coupling regime with $\theta = \pi/5$ and energy loss $\alpha L_b = 0.008$, where the photonic bands are topologically trivial. As the coupling between sites in increased (without changing the energy loss) to $\theta = 0.32\pi$, nontrivial edge modes (orange and blue dots) span the topological band gap highlighted in the blue area in the middle panel of Fig. 2(b). If the coupling $\theta$ (which depends only on the wavelength of light and geometric parameters) is held constant at $\theta = 0.32\pi$ but the energy loss is further increased, the effective coupling between the lattice sites decreases, eventually closing the topological band gap at the phase transition and reopening a trivial bulk gap in the heavily lossy phase, as shown in the right panel for $\alpha L_b = 0.92$.

The imaginary part of $\phi(k)$ in the topological phase for $\theta = 0.32\pi$ and $\alpha L_b = 0.008$ is plotted in Fig. 2(c). $\Im[\phi(k)]$ corresponds to the decay rate of the photons due to dissipation for either pesudospin, and is momentum independent, as shown in the left panel. The imaginary part of $\phi$ scales linearly with the energy loss $\alpha L_b$, as indicated in the panel on the right.

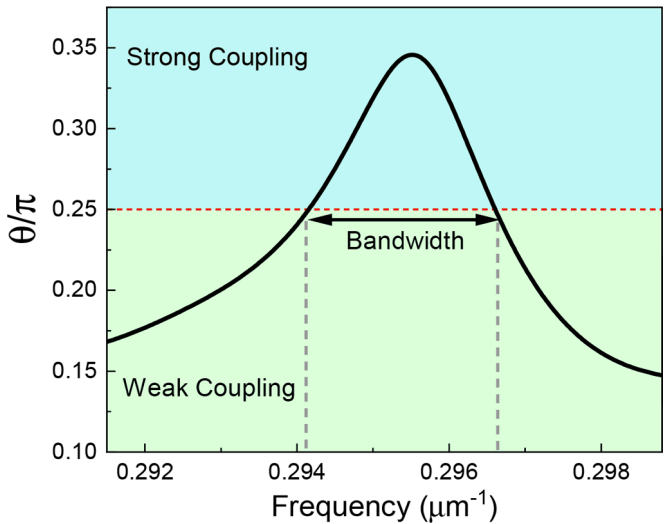

FIG. 3. Site ring coupling strength as a function of frequency, showing the strong coupling bandwidth. By tuning the input laser wavelength within the topological spectral bandwidth, the proposed sensor can probe the spectral lines of the chemical species of interest, permitting adsorption spectroscopy.

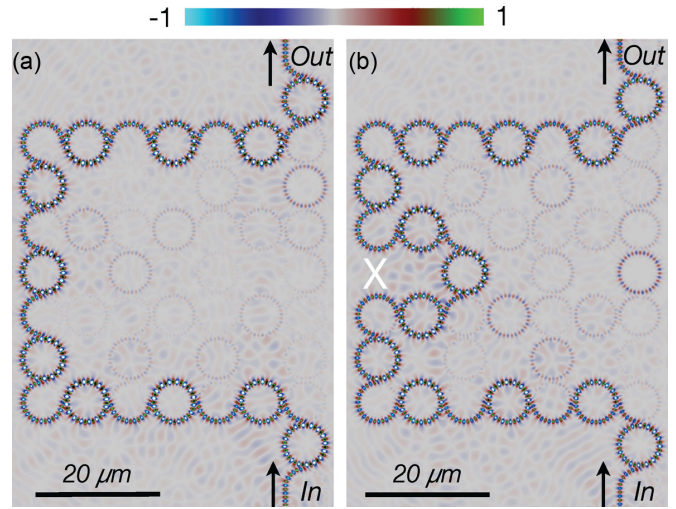

FIG. 4. FDTD simulation for a topological photonic array. SWG waveguides operate at $\lambda = 3.4$ $\mu$m and are exposed to a methane concentration of 1%, with energy loss 0.0022. The colors indicate the intensity of the electric field in the lattice. (a) Topologically protected edge states along the lattice boundary; (b) in the presence of a defect at the boundary, the topological edge state reroutes around the perturbation, preserving the transmission of light across the array.

As it will be clear in the next section, for trace gas concentrations ($\sigma < 1\%$), the energy loss is very small $\alpha L_b \ll 0.1$. In the strong coupling regime $\theta > \pi/4$, except for couplings that are in the near vicinity of the critical coupling, $\theta \sim \pi/4$, typical absorption losses are not expected to suppress topological protection, as indicated in Fig. 2(a). For applications that require very large loss, however, topological protection could be in principle suppressed even in the absence of backscattering. As promised, we address in the next section realistic simulations that account for backscattering effects, including array defects and dislocations. Those results confirm the analysis based on the scattering matrix theory in the absence of intrinsic disorder. They also indicate that the topological modes remain fairly robust even in the presence of a significant amount of disorder.

## IV. ABSORPTION SPECTROSCOPY SIMULATION

In topologically protected photonic lattices, light energy injected into the input port will flow along the lattice edge to the output port, as illustrated in Fig. 1(c). Thus the underlying sensing mechanism consists of monitoring the transmission loss due to light absorption by a chemical species [8,26].

The photonic topological insulator is the $3 \times 3$ lattice shown in Fig. 3. The site rings are given the radius $R_s = 4.074$ $\mu$m, whereas the bridge resonators have the radius $R_b = 4.2$ $\mu$m, with $Q = 10^4$ and the spacing between two rings set at $g = 0.3$ $\mu$m. We perform finite domain time difference (FDTD) simulations to demonstrate a topological sensor operating at the wavelength $\lambda = 3.4$ $\mu$m, targeting methane's mid-IR absorption band. The corresponding molar absorption for methane is $\epsilon = 8.21$ cm$^{-1}$ atm$^{-1}$[31]. We note that if we take these geometric parameters and methane gas as an absorbing medium, from Eq. (4), we find that even for a methane concentration $\sigma = 100\%$ the maximum value of energy loss is $\alpha L_b \sim 10^{-2}$–$10^{-3}$ for $m \in (10$–$100)$. Thus, for reasonable values of the structural parameters, gas absorption (and particularly trace gas absorption) is not enough to break topological protection.

We adopt a SWG waveguide of segmented silicon in air cladding, with refractive indexes $n_1 = 3.4$ and $n_2 = 1$, respectively. The SWG has spacing $\Lambda = 0.2$ $\mu$m, thickness $a = 0.12$ $\mu$m, and height $w = 0.5$ $\mu$m. The coupling strength between two site rings is found to be $\theta = 0.32\pi$ and $\phi(\lambda = 3.4$ $\mu$m$) = 0.24\pi$. Those parameters fall in the range were topological protection is expected according to Fig. 2(a). The FDTD simulations below (explained in detail in Appendix A) are fully consistent with the topological character predicted for the edge modes. The simulation results summarized in Fig. 3 show that the topological bandwidth is in the range 2940–2966 cm$^{-1}$. Hence, the spectral bandwidth is $\Delta\lambda \sim 26$ cm$^{-1}$ inside the topological gap. This feature permits tuning the input laser wavelength to differentiate the spectral structure of methane's fine fingerprint pattern, enabling TDLAS.

In Fig. 4, we show the electric field diagrams as obtained from the FDTD simulations for a methane concentration $\sigma$ of 1%, which corresponds to an energy loss $\alpha L_b = 0.0022$. The additional resonator added to the output port forces the transmitting energy to funnel into the output channel rather than loop back into the lattice. Figure 4(a) demonstrates the propagation of topological edge modes in the presence of dissipation, in agreement with the results of the scattering matrix theory. In Fig. 4(b), the lattice is dented by removing a bridge resonator from the edge. Defects are commonly introduced by dislocations formed in the fabrication process, which are described in more detail in Appendix C. As expected, the photonic edge states are topologically robust and accordingly reroute around this perturbation. As shown in Appendix C, the sensing device is also robust against spatial fluctuations in the size of the resonators due to typical fabrication resolution effects [32].

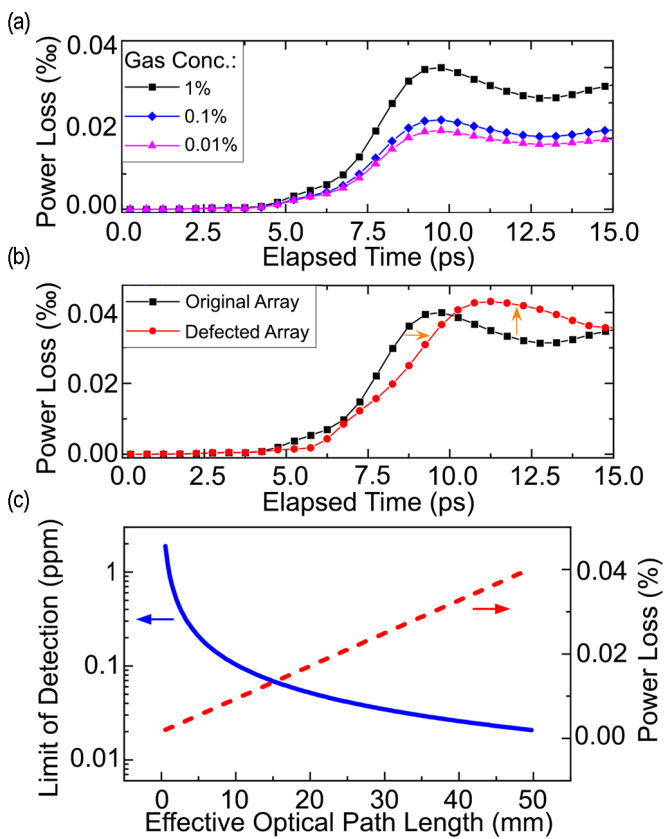

FIG. 5. Time dependent power loss simulation of the topological gas sensing performance: (a) under different trace gas concentrations and (b) with and without a defect for a methane concentration of 1%. (c) Predicted limit of detection (blue) in ppm vs the effective optical path length $L$ along the edge. The detection limit scales as $1/L$. The red curve shows the total power loss. It scales linearly with $L$, for a fixed concentration $\sigma$ of 0.1%.

The proposed absorption-based topological sensor monitors the transmission loss along the edge states in real time. The time-dependent transmission results across the photonic lattice are shown in Fig. 5(a). The power absorbed over time is scanned for trace methane concentrations down to 0.01%. As shown in the plots, it takes ∼10 picoseconds for the power absorption to stabilize. In Fig. 5(b), we compare the power absorption of the topological photonic insulator with and without disorder at the edge [Figs. 4(a) and 4(b), respectively]. The longer path with the disorder leads to an proportionally increased power loss and a correspondingly longer time delay for the steady state transmission, indicated by the arrows. By scaling up the system size and increasing the optical path length by orders of magnitude, a well-designed TDLAS based topological sensor can possibly reach very high levels of chemical sensitivity, in spite of fabrication disorder effects.

Depending on the application for which gas sensing is needed, an appropriate detection range might vary from thousands of parts per million (ppm) [33] down to parts per billion (ppb) range [34,35]. In the topological phase, the power attenuation scales exponentially with the combination $\sigma L$, with $L$ the total optical path length along the edge. For a fixed small energy loss ($\sigma L \ll 1/\epsilon\gamma$), the concentration hence scales with the length of the edges as $\sigma \propto 1/L$. If one fixes the energy loss at the limit of the sensitivity, which is set by the laser power, the detection limit should therefore scale linearly with the inverse of the system size in the topological regime. This behavior is consistent with the scaling analysis of the simulation results performed for a fixed energy loss at different system sizes, as shown in Appendix B.

The plot with the sensor's limit of detection as a function of the sensing path length [blue curve in Fig. 4(c)] demonstrates that sensitivities below 1 ppm can be easily reached by an optical path length $L$ of less than 1 cm. That sensing length scale corresponds to a photonic lattice of $43 \times 43$ sites. In addition, the detection limit below 100 ppb requires topologically protected sensing paths of ∼5 cm length, corresponding to a lattice of $215 \times 215$ sites. The red curve is total power absorbed by the gas medium for a fixed concentration of methane ($\sigma = 0.1\%$) and scales linearly with $L$. The sensing performance of this topological device could be further improved by optimizing the interaction factor $\gamma$ of the SWG and the $Q$-factor of the lattice of resonators.

## V. CONCLUSION

Our midinfrared numerical simulations demonstrate that topological photonic sensors permit an efficient and reliable detection of trace amounts of a gas irrespective of most kinds of disorder and environmental noise. We point out that this fundamentally new design strategy can serve as a building block to construct more sophisticated on-chip architectures. Photonic crystal waveguides with large footprints have been widely used in integrated devices [36,37]. We propose that absorption spectroscopy utilizing topological edge states could empower the development of novel integrated photonic sensing technologies for on-chip chemical detection applications.

### ACKNOWLEDGMENTS

Authors thank M. Santos for useful discussions. B.U. acknowledges Carl T. Bush fellowship and National Science Foundation Grant DMR-2024864 for the partial support. Author Y.Z. acknowledges National Natural Science Foundation of China Grant 61705099 and Natural Science Foundation of Shanghai Grant 21ZR1443100 for the partial support. Author B.W. acknowledges OU Big Idea Challenge award for the partial support.

## APPENDIX A: NUMERICAL STUDY OF THE STRONG COUPLING CONDITION

The FDTD simulations were performed using Rsoft Photonics Version 2019.09 via the FULLWAVE tool. The materials used in the simulation are silicon and methane. Perfectly matched layers are used on the boundaries of the simulation to prevent reflections. The largest mesh element size is set lower than 1/100 of the lowest wavelength.

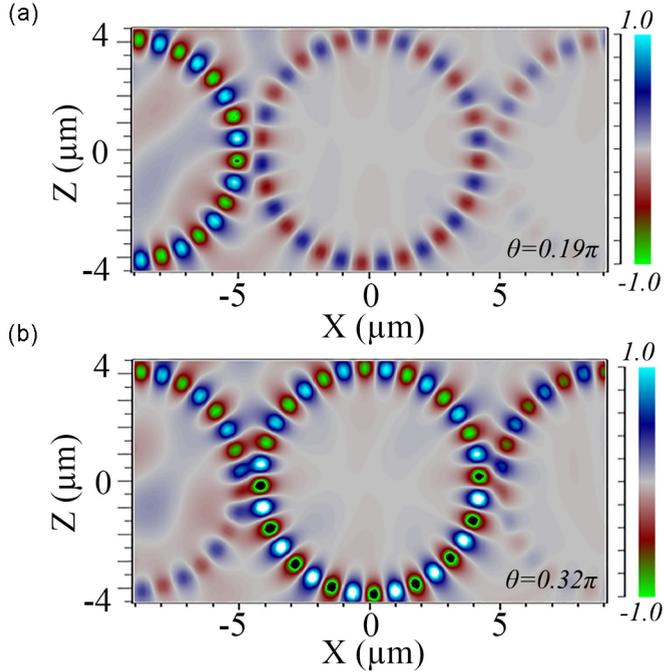

FIG. 6. FDTD simulation for $\theta = 0.19\pi$ (top) and $0.32\pi$ (bottom).

The sensitivity and limit of detection of TDLAS in the photonic lattice is modelled using realistic system configurations [26] that include the laser injecting power and photodiode's detection limit [8,26]. Using a waveguide input power of 5 mW, the sensor's sensitivity and limit of detection can be obtained with an estimation of the photodiode's minimum detectable power. The minimum detectable power is given in terms of the photodiode signal-to-noise ratio (SNR), noise equivalent power (NEP), and bandwidth (BW) by $\delta P_{\min} = \text{SNR} \times \text{NEP} \times (\text{BW})^{\frac{1}{2}}$. A commercial mid-IR InAs photovoltaic detector (such as the Hamamatsu P10090 series) can achieve an NEP of $1.5 \times 10^{-13}$ W/$\sqrt{\text{Hz}}$. Thus the configuration offers about 60-dB power budget to the system. If the detection bandwidth BW is limited to 2 MHz with a signal to noise ratio of 10, an estimated theoretical limit of detection (LoD) can be obtained for each sensitivity.

The determination of the coupling $\theta$ between site rings can be done via FDTD using the relation: $\theta = \sqrt{I_{\text{out}}/I_{\text{in}}}$ where $I_{\text{out}}$ and $I_{\text{in}}$ are the intensities in the input and output ports. In Fig. 6, we show the field amplitudes for $\theta = 0.19\pi$, in the topologically trivial regime, and $\theta = 0.32\pi$ in the topologically nontrivial one. Further, by scanning the ring resonator's transmission in the frequency spectrum, a region of strong coupling behavior that enables the topological effect can be obtained. Figure 3 in the main text presents the spectral bandwidth in the topological sector of the phase diagram.

## APPENDIX B: SCALING OF THE DETECTION LIMIT

In order to extract the detection limit for large system sizes, we performed simulations in five different system sizes ranging from $3 \times 3$ to $3 \times 7$ lattices and scaled the data. The data points for limit of detection versus effective optical path length are shown in Table I and were plotted in Fig. 7. The scaling of the data with the inverse of the optical path length $1/L$ shown in the same plot is a straight line passing through the origin.

TABLE I. Power absorbed in the photonic topological insulator and the sensing limit of detection (LoD) when the length of the effective optical path is varied by scanning the lattice size from $3 \times 3$ to $3 \times 7$.

| Effective Optical Path (mm) | Power Absorbed (a.u.) | Limit of Detection (ppm) | Lattice size |
|---|---|---|---|
| 0.567 | 0.00206 | 1.823 | $3 \times 3$ |
| 0.671 | 0.00215 | 1.54 | $3 \times 4$ |
| 0.775 | 0.00219 | 1.33 | $3 \times 5$ |
| 0.879 | 0.0023 | 1.18 | $3 \times 6$ |
| 0.983 | 0.00239 | 1.05 | $3 \times 7$ |

## APPENDIX C: DISORDER AND DEFECTS INTRODUCED BY FABRICATION IMPERFECTIONS

Under realistic fabrication conditions, disorder is introduced in the array of optical resonators by intrinsic fabrication resolution limits. The writing resolution defines the minimum feature size that the electron-beam lithography (EBL) technique can produce. Disorder and defects are also introduced by the stitching process of the EBL writing tool [32,38,39]. The stitching accuracy corresponds to errors introduced in the process of connecting writing patches together due to the limited one-time maximum writing area that EBL provides. Below we incorporate those two intrinsic disorder factors in the FDTD simulation and explicitly compare the performance of topologically protected 2D arrays of optical resonators with a trivial 1D chain of resonators.

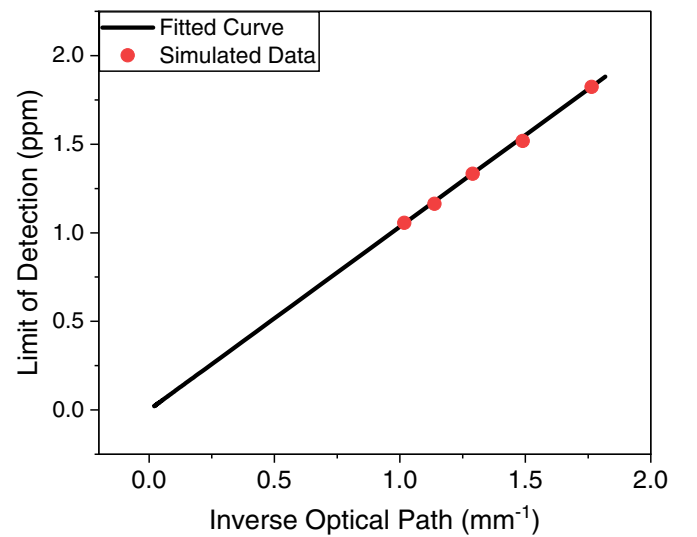

FIG. 7. Scaling of the simulation data for the limit of sensitivity for different system sizes. The scaling shows that the limit of sensitivity scales linearly with the inverse of the system size.

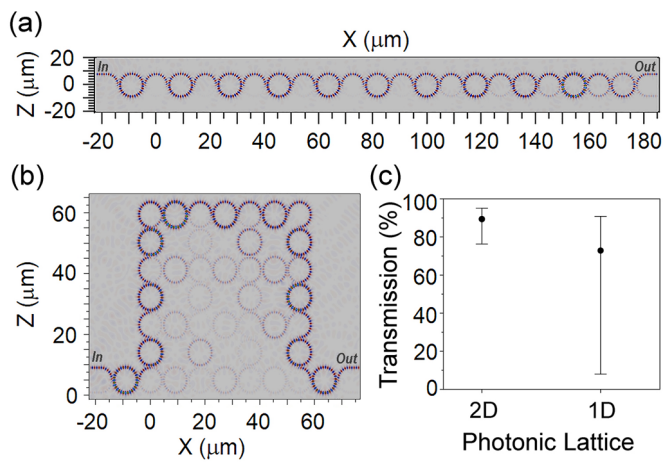

FIG. 8. (a) FDTD simulation of a disordered 1D photonic array of resonators with local fluctuations $\delta R \in [0, \pm 25]$ nm randomly assigned to each ring. (b) FDTD simulation the same type of disorder ($\delta R \in [0, \pm 25]$ nm) in a 2D topological array of resonators. In both simulations, light is transmitted from the left to the right of the lattice. The topological edge states are robust against the disorder within a resolution tolerance of 50 nm. (c) Power transmission performance averaged over five different disorder realizations in the 2D topological array and in the 1D trivial one. The transmission in the 1D array is much less stable, depending on favorable configurations of the disorder even in short chains.

The material loss due to absorption in silicon is very low and typically less than 1 dB/cm. For a power budget of 60 dB under a 2 MHz detection bandwidth (assuming an input laser power is of 5 mW and the noise equivalent power (NEP) of $1.5 \times 10^{-13}$ W/$\sqrt{\text{Hz}}$ from a commercial InAs detector), that loss would permit scaling the system up to a very large array of $4000 \times 4000$ resonators (with a total optical path length of 60 cm) before the transmission output would be comparable to the noise at the detector.

#### 1. Resolution limit impact

EBL is the most flexible and straightforward way to pattern photonic waveguides with a critical dimension of $\sim$10 nm. The method uses a focused beam of electrons to draw shapes on a substrate coated with an electron-sensitive film called a resist. With EBL, one can typically achieve 10-nm resolution. The resolution limit is determined by the spot size of the focused beam and is influenced by forward scattering of the e-beam inside the resist and by backscattering from the substrate.

To verify the robustness of the topological edge states to intrinsic disorder effects due to the resolution of the writing tool, we introduced random fluctuations in the size of individual resonator rings $\Delta R$, where $R(x) = R + \delta R(x)$, and simulated the transmitance averaged over several disorder realizations. Our simulations show that the topological protection in the 2D array proposed in the manuscript is maintained within a resolution tolerance up to 50 nm per ring [e.g., $\delta R(x)$ a random number between 0 to $\pm 25$ nm for each site], with the edge states largely unaffected.

In Fig. 8(b), we show the simulation of a 2D topological design for a patterning resolution of 50 nm, which is considerably poorer than the EBL's best resolution at 10 nm [32]. Even in this case, there is a robust transmission of light through the topological edge states of the 2D lattice. We averaged the disorder over five different realizations. The averaged power transmission [shown in Fig. 8(c) is at 90%]. Although the simulation is performed in a $3 \times 3$ array, the system can be easily scaled up. When offered a 60 dB power budget to the sensing chip, the system can be increased up to an array of $390 \times 390$ resonators before the transmission output becomes comparable to the detector noise. This simulation demonstrates that the proposed topological design is robust and can be realistically realized.

For comparison, a similar simulation with a trivial chain of resonators [shown in Fig. 8(a)] shows an average transmittance 17% lower and a much less stable power transmission even for relatively short chains, as indicated by the much broader range of variation in the power transmission [see Fig. 8(c)]. The average transmission deteriorates quickly with increasing the system size, as it relies on favorable (and rare) configurations of the disorder. The largest drawback of the trivial 1D design, compared to the topological one, is its extreme susceptibility to failure due to defects introduced by the stitching process of the EBL tool, which are addressed in the next subsection.

#### 2. Stitching accuracy impact

Devices with a high sensitivity require long optical path lengths, which can be achieved in relatively large photonic sensing chips. Due to the fact that EBL has a limited area of the one-time writing field, the whole photonic array will extend over a series of EBL writing fields that are stitched together by the lithography control system. The current state-of-art EBL systems have a stitching accuracy at $\pm 10$ nm.

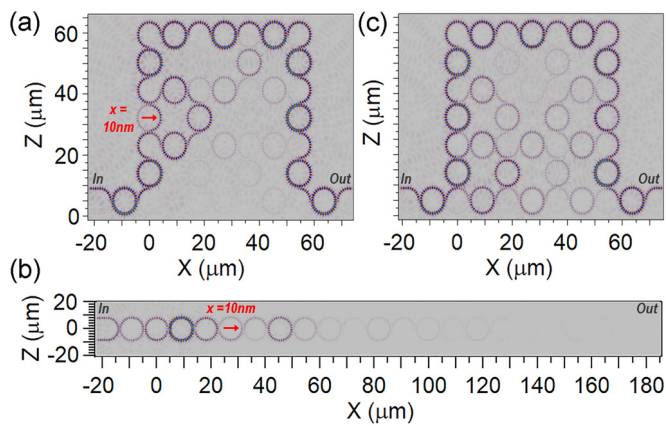

FIG. 9. FDTD simulations of (a) the 2D topological array of resonators and (b) a 1D trivial array of resonators, with a 10-nm stitching error shifting one ring in $x$ direction, as indicated by the red arrow. (c) FDTD simulation of the disordered 2D topological array, in which every ring contains a randomly introduced dislocation (stitching error) along the $x$ and $y$ directions ($\delta x, \delta y$), varying from 0 to $\pm 10$ nm. The edge states can easily reroute around dislocation induced defects, whereas in the trivial 1D chain, they completely suppress transmittance.

The stitching accuracy typically introduces geometric distortions, e.g., position shifting errors, which can potentially attenuate or even block the light transmission from one subpattern to its adjacent one [38]. Those errors in the position of the rings can introduce defects in the form of dislocations in the lattice.

To incorporate the stitching errors in the FDTD simulation, we introduce a random pattern of dislocations which offset the position of individual rings. While the EBL writing field can be configured in the range of $1 \times 1$ $\mu m^2$ to $100 \times 100$ $\mu m^2$ (e.g., JOEL-JBX, STS-ELS series), for simplicity, we choose a writing field of $10 \times 10$ $\mu m^2$. In this case, we can simply fit one ring in each writing field.

The effect of the dislocation of a single ring by 10 nm is shown in the FDTD simulation for the cases of a topological 2D array [Fig. 9(a)] and in a trivial 1D chain [Fig. 9(b)]. In the former, the topological edge states reroute around the dislocation, which effectively behaves as a defect. In that case, 80% of the power is transmitted to the output. In the trivial 1D chain, the dislocation also behaves as a defect, but the transmission is nearly completely blocked. Similar results are found regardless of the direction of the dislocation. In Fig. 9(c), we consider a more realistic situation with a random distribution of dislocations (stitching error) at every site. We assume dislocations in the $x$ and $y$ directions, with $\delta x$ and $\delta y$ varying at each site in the range from 0 to ±10 nm. The simulation shows that the 2D topological system can operate normally with defects introduced by dislocations inherent to the stitching process. In the trivial 1D system, defects introduced by dislocations will completely disrupt transmittance.

[1] Green William, Chi Xiong, Marwan Khater, Yves Martin, Eric Zhang, Chu Teng, Jason Orcutt *et al.*, *Methane trace-gas sensing enabled by silicon photonic integration*, Tech. Rep. No. DOE-IBM-0000540-8 (IBM Research, University of Oklahoma, Norman, Oklahoma, 2017).

[2] A. R. Brandt, G. A. Heath, E. A. Kort, F. O'Sullivan, G. Pétron, S. M. Jordaan, P. Tans, J. Wilcox, A. M. Gopstein, D. Arent, *et al.*, Methane leaks from north american natural gas systems, Science **343**, 733 (2014).

[3] B. P. J. de Lacy Costello, M. Ledochowski, and N. M. Ratcliffe, The importance of methane breath testing: a review, J. Breath Res. **7**, 024001 (2013).

[4] M. C. Estevez, M. Alvarez, and L. M. Lechuga, Integrated optical devices for lab-on-a-chip biosensing applications, Laser Photonics Rev. **6**, 463 (2012).

[5] M. Lackner, Tunable diode laser absorption spectroscopy (td-las) in the process industries–a review, Rev. Chem. Eng **23**, 65 (2007).

[6] N. M. M. Pires, T. Dong, U. Hanke, and N. Hoivik, Recent developments in optical detection technologies in lab-on-a-chip devices for biosensing applications, Sensors **14**, 15458 (2014).

[7] V. Passaro, C. D. Tullio, B. Troia, M. L. Notte, G. Giannoccaro, and F. D. Leonardis, Recent advances in integrated photonic sensors, Sensors **12**, 15558 (2012).

[8] A. Gervais, P. Jean, W. Shi, and S. LaRochelle, Design of slow-light subwavelength grating waveguides for enhanced on-chip methane sensing by absorption spectroscopy, IEEE J. Sel. Top. Quantum Electron. **25**, 5200308 (2018).

[9] V. Van, *Optical Microring Resonators: Theory, Techniques, and Applications* (CRC Press, University of Oklahoma, Norman, Oklahoma, 2016).

[10] F. D. M. Haldane and S. Raghu, Possible Realization of Directional Optical Waveguides in Photonic Crystals with Broken Time-Reversal Symmetry, Phys. Rev. Lett. **100**, 013904 (2008).

[11] Z. Wang, Y. D. Chong, J. D. Joannopoulos, and M. Soljačić, Reflection-Free One-Way Edge Modes in A Gyromagnetic Photonic Crystal, Phys. Rev. Lett. **100**, 013905 (2008).

[12] T. Ozawa, H. M. Price, A. Amo, N. Goldman, M. Hafezi, L. Lu, M. C. Rechtsman, D. Schuster, J. Simon, O. Zilberberg *et al.*, Topological photonics, Rev. Mod. Phys. **91**, 015006 (2019).

[13] A. B. Khanikaev, S. H. Mousavi, W.-K. Tse, M. Kargarian, A. H. MacDonald, and G. Shvets, Photonic topological insulators, Nat. Mater. **12**, 233 (2013).

[14] T. Ochiai, Photonic realization of the (2+ 1)-dimensional parity anomaly, Phys. Rev. B **86**, 075152 (2012).

[15] M. C. Rechtsman, J. M. Zeuner, Y. Plotnik, Y. Lumer, D. Podolsky, F. Dreisow, S. Nolte, M. Segev, and A. Szameit, Photonic floquet topological insulators, Nature (London) **496**, 196 (2013).

[16] M. Hafezi, E. A. Demler, M. D. Lukin, and J. M. Taylor, Robust optical delay lines with topological protection, Nat. Phys. **7**, 907 (2011).

[17] G. Q. Liang and Y. D. Chong, Optical Resonator Analog of A Two-Dimensional Topological Insulator, Phys. Rev. Lett. **110**, 203904 (2013).

[18] F. D. M. Haldane, Model for A Quantum Hall Effect without Landau Levels: Condensed-Matter Realization of the" Parity Anomaly", Phys. Rev. Lett. **61**, 2015 (1988).

[19] C. L. Kane and E. J. Mele, Quantum Spin Hall Effect in Graphene, Phys. Rev. Lett. **95**, 226801 (2005).

[20] S. Mittal, E. A. Goldschmidt, and M. Hafezi, A topological source of quantum light, Nature (London) **561**, 502 (2018).

[21] M. A. Bandres, S. Wittek, G. Harari, M. Parto, J. Ren, M. Segev, D. N. Christodoulides, and M. Khajavikhan, Topological insulator laser: Experiments, Science **359**, eaar4005 (2018).

[22] J. E. Heebner, V. Wong, A. Schweinsberg, R. W. Boyd, and D. J. Jackson, Optical transmission characteristics of fiber ring resonators, IEEE J. Quantum Electron. **40**, 726 (2004).

[23] F. Liu, Q. Li, Z. Zhang, M. Qiu, and Y. Su, Optically tunable delay line in silicon microring resonator based on thermal nonlinear effect, IEEE J. Sel. Top. Quantum Electron. **14**, 706 (2008).

[24] P. Cheben, R. Halir, J. H. Schmid, H. A. Atwater, and D. R. Smith, Subwavelength integrated photonics, Nature (London) **560**, 565 (2018).

[25] H. Kikuta, H. Yoshida, and K. Iwata, Ability and limitation of effective medium theory for subwavelength gratings, Optical Rev. **2**, 92 (1995).

[26] A. Gutierrez-Arroyo, E. Baudet, L. Bodiou, V. Nazabal, E. Rinnert, K. Michel, B. Bureau, F. Colas, and J. Charrier, Theoretical study of an evanescent optical integrated sensor for multipurpose detection of gases and liquids in the mid-infrared, Sens. Actuators B Chem. **242**, 842 (2017).

[27] A. Nitkowski, L. Chen, and M. Lipson, Cavity-enhanced on-chip absorption spectroscopy using microring resonators, Opt. Express **16**, 11930 (2008).

[28] M. S. Luchansky and R. C. Bailey, High-q optical sensors for chemical and biological analysis, Anal. Chem. **84**, 793 (2012).

[29] M. Pasek and Y. D. Chong, Network models of photonic floquet topological insulators, Phys. Rev. B **89**, 075113 (2014).

[30] F. Gao, Z. Gao, X. Shi, Z. Yang, X. Lin, H. Xu, J. D. Joannopoulos, M. Soljačić, H. Chen, L. Lu, *et al.*, Probing topological protection using a designer surface plasmon structure, Nat. Commun. **7**, 11619 (2016).

[31] H. Tai, H. Tanaka, and T. Yoshino, Fiber-optic evanescent-wave methane-gas sensor using optical absorption for the 3.392-$\mu$m line of a he–ne laser, Opt. Lett. **12**, 437 (1987).

[32] A. N. Broers, A. C. F. Hoole, and J. M. Ryan, Electron beam lithography—resolution limits, Microelectron. Eng. **32**, 131 (1996).

[33] X. Jia, J. Roels, R. Baets, and G. Roelkens, On-chip non-dispersive infrared co2 sensor based on an integrating cylinder, Sensors **19**, 4260 (2019).

[34] O. Casals, N. Markiewicz, C. Fabrega, I. Gràcia, C. Cané, H. S. Wasisto, A. Waag, and J. D. Prades, A parts per billion (ppb) sensor for no2 with microwatt ($\mu$w) power requirements based on micro light plates, ACS Sens. **4**, 822 (2019).

[35] T. Tomberg, M. Vainio, T. Hieta, and L. Halonen, Sub-parts-per-trillion level sensitivity in trace gas detection by cantilever-enhanced photo-acoustic spectroscopy, Sci. Rep. **8**, 1848 (2018).

[36] A. Salmanpour, S. Mohammadnejad, and A. Bahrami, Photonic crystal logic gates: an overview, Opt. Quantum Electron. **47**, 2249 (2015).

[37] Y. Zhang, Y. Zhao, and R.-qing Lv, A review for optical sensors based on photonic crystal cavities, Sensors and Actuators A: Physical **233**, 374 (2015).

[38] A. L. Bogdanov, J. Lapointe, and J. H. Schmid, Electron-beam lithography for photonic waveguide fabrication: Measurement of the effect of field stitching errors on optical performance and evaluation of a new compensation method, J. Vac. Sci. Technol. B **30**, 031606 (2012).

[39] Y. Lin, B. Yu, Y. Zou, Z. Li, C. J. Alpert, and D. Z. Pan, Stitch aware detailed placement for multiple e-beam lithography, Integration **58**, 47 (2017).